\documentclass[aps,pra,showpacs,twocolumn]{revtex4-1}
\usepackage{graphicx,amsmath,amssymb}
\usepackage{silence}
\usepackage[usenames]{color}
\usepackage{mathrsfs}
\usepackage{indentfirst}
\usepackage{float}
\usepackage{braket}
\usepackage[colorlinks=true, citecolor=blue, urlcolor=blue, linkcolor=blue ]{hyperref}
\usepackage{epstopdf}
\hypersetup{breaklinks=true}
\begin{document}
	
	\title{Temperature-induced dephasing in high-order harmonic generation from solids}
	
	\author{Tao-Yuan Du }\email{Corresponding author. duty710@163.com}  \affiliation{School of Mathematics and Physics, China University of Geosciences, Wuhan 430074, China} % kai,hei,song,gbsn,gkai,bmsi,bkai
	\author{Chao Ma}   \affiliation{School of Mathematics and Physics, China University of Geosciences, Wuhan 430074, China}

	% \date{\today}% It is always \today, today,
	
	\begin{abstract}
		High harmonic generation (HHG) in solid and gaseous targets has been proven to be a powerful avenue for the generation of attosecond pulses, whereas the influence of electron-phonon scattering on HHG is a critical outstanding problem. Here we first introduce a temperature dependent lattice vibration model by characterizing the spacing fluctuation. Our results reveal that (i) structural disorder induced by lattice vibration does not lead to generation of even-order harmonics; (ii) dephasing of HHG occurs as the lattice temperature is growing; (iii) an open-trajectory picture predicts the maximal photon energy in the temperature-dependent HHG spectra. Moreover, a formula assessing dephasing time with lattice temperature is proposed to identify the timescale of electron-phonon scattering. This work paves a way to study non-Born-Oppenheimer effect in solids driven by strong field.
	\end{abstract}
	
	\maketitle
	
	\section{Introduction}
	High harmonic generation (HHG) from solids is at the forefront of an ongoing attosecond science and strong field physics \cite{Chin,Krausz}. Solid-state HHG provides a significant view to explore the emerging condensed matter systems \cite{solid1,solid2,solid3,solid4,Du3,Du4,solid5,solid6,solid7,solid8,Du4,Du2}, for example, two-dimensional materials \cite{graphene1,graphene2}, topological insulators \cite{topo1,topo2}, mott insulators \cite{mott1,mott2}, disorder solids \cite{disorder1,disorder2,disorder3,disorder4} and liquids \cite{luu}. 	
	In contrast to HHG in gases, strong-field induced electronic dynamics among the periodic nuclei inevitably suffers the many-body scatterings. Moreover, the high intensity and repetition rate of strong laser pulses will lead to lattice thermal effect. An open question is the failure to understand the significant discrepancy between experimental HHG spectra and theoretical those, where the latter requires an extremely short dephasing times \cite{solid4,solid5,solid6}.
	
	The theoretical HHG simulations are lack of the classification on many-body interactions, it is challenging to finely understand the solid-state HHG experiments. Currently, all of many-body interactions in solid-state HHG are included wholly by an indistinguishable dephasing parameter. One of critical many-body interactions is the scattering caused by phonon (i.e., quasi particles reflecting quanta of lattice vibration).

	The non-Born-Oppenheimer full ab-initio quantum treatments of condensed matter system are a nearly impossible task due to the limit of available computing resource. 
	Therefore, a new lattice model involving the temperature-dependent nuclear vibration is desirable to unravel the non-Born-Oppenheimer effect of nuclei in solid-state HHG.
	Less is known about the temperature-dependent nuclear vibration in the  experimental and theoretical HHG from solids, although several results of temperature-dependent HHG from solids \cite{Uchida} and liquids \cite{Heissler} displayed the sharply decreasing HHG yields with the increasing temperature. Actually, those results imply dephasing but are lack of explicit discussions.

	In this work, we develop a finite-temperature crystal model characterizing the temperature-dependent distribution of nuclei to capture one of real-space lattice vibration modes (i.e. structural configurations), and propose a formula of extracting dephasing time that can agree perfectly with the "experimental" observations. It opens up the possibilities to study the temperature-dependent quantum decoherence in solid-state high-harmonic spectroscopy.

\section{Model}
\begin{figure}[t]	\centering	\includegraphics[width=8.5 cm, height= 8 cm]{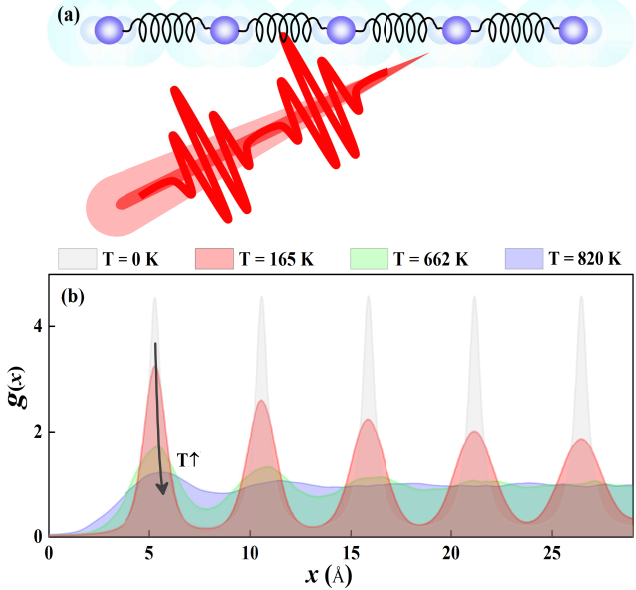}
		\caption{(a) Schematic diagram of the interaction between the femtosecond laser pulses and the vibrating atomic chain. The light-color balls around each bright ball denote the deviation from the equilibrium position. (b) The radial distribution function with varying lattice temperatures. }
		\label{Fig1}
	\end{figure}

	To investigate the impact of lattice temperature on HHG, we first propose a model to assess the variation of nuclear location under the different temperatures. 
	In Fig. \ref{Fig1}(a) the real-space lattice-vibration modes at specific temperature appear as the disordered atomic positions $x_j$ ($j$ = $1,2,...,N$) and can be approximately modeled by a Gaussian probability distribution function. Nuclear spacing $\xi_j$ = $x_{j+1} - x_{j}$ for each vibration mode is assumed to obey the standard normal distribution $f(\xi)$ = $(1/\sqrt{2\pi \sigma^2})$exp\{-[($\xi  - a)^2/2\sigma^2$)]\} with the mean spacing value i. e. lattice constant $a$ = 10 a.u. (atomic units are used throughout unless otherwise indicated) and variance ${\sigma ^2}$. The fixed-nuclei chain ($\sigma$ = 0) is a periodic arrangement with lattice constant $a$. However, nuclei will deviate from their equilibrium positions under the non-Born-Oppenheimer picture. 
	Since the coulomb repulsion extremely increases as two nuclei approach each other, intervals between nuclei have a lower limit. Similarly, an upper limit of the nuclear intervals is also necessary before the structural damage. For the rationality of our model, a truncated normal distribution generator is adopted and thus only the atomic-pair intervals falling within a symmetric range $\left[a-\zeta, a+\zeta \right]$ are retained \cite{JMziman,RDF1}. Here $\zeta$ adopted $a$/3 in simulations is the maximal deviation under the harmonic oscillator model.

Based on the Debye model \cite{Debye,RDF2}, we introduce a temperature dependent lattice vibration model, in which one refers to the details in Appendix A. The expression of structure fluctuation under certain temperature can be written as
	\begin{equation}\label{Eq1}
		\sigma^2 = \frac{{6}}{{M{w_D}}}\left[ {\frac{1}{4} + {{\left( {\frac{\text{T}}{{{\Theta _D}}}} \right)}^2}\Phi } \right],
	\end{equation}
where $M$ is the atomic mass, ${\Theta _D} = \hbar {w_D}/{k_B}$ is the Debye temperature, and $\Phi  = \int _0^{{\Theta _D}/T}k{({e^k} - 1)^{ - 1}}dk$ involves the contribution from the all phonon modes described by phonon density of states. Keep the truncation in mind, the corresponding relation between fluctuation $\sigma$ (or the effective fluctuation $\sigma^{*}$) and lattice temperature is clarified in Appendix B.

The radial distribution function (RDF), an experimental order parameter, is the number of atoms in the shell between x and x + dx. Counting the distance ${\xi _j}$ between any atomic pairs, RDF is expressed as $\text{g}(x) = \sum_j {\delta (x - {\xi _j})}$. The structure of RDF could characterize the variation of atomic-pair distributions caused by growing lattice temperatures \cite{x-ray,neutron,electron}. As shown in Fig. \ref{Fig1}(b), the amplitudes of long-range peaks decrease faster than those of the short-range peaks. These characters are in good agreement with the experiments, in which the effect of temperature on RDF had been studied  \cite{Urquidi,Ojovan}. For each structural configuration, the nuclear spacing arrangement displays the destruction of the long-range order, but the short-range order is still kept well before the damage of atomic structure. To achieve our lattice-vibration model with the solid characteristic of periodic translation invariant, in structural configurations we extremely reduce the level of atomic-chain disorder via including the mirror symmetry about the nucleus-vibrating equilibrium position. This scheme includes the phase-breaking events reflecting the electron-phonon scattering. The inclusion of periodic arrangement and phase breaking determines intrinsically the uniqueness of our model \cite{zengaiwu}.

	\begin{figure}[t]
		\includegraphics[width=9 cm,height=8 cm]{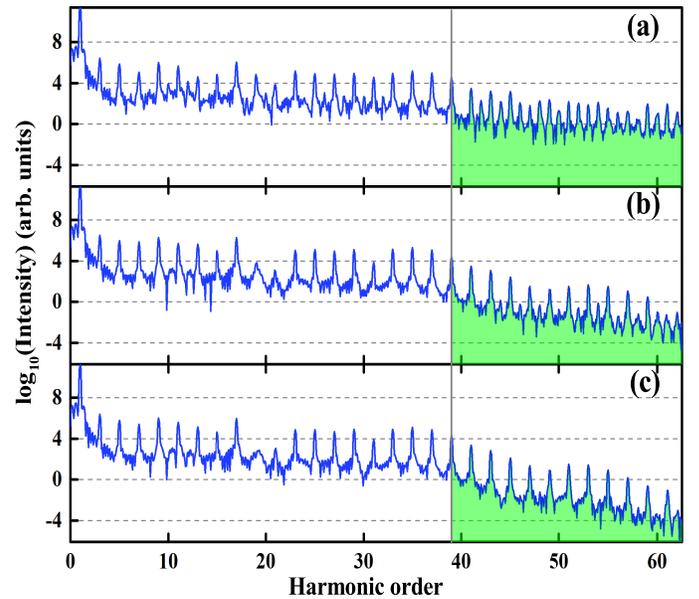}
		\caption{(a), (b) and (c) respectively present HHG spectra obtained from 1, 100 and 4000 configurations, in which the lattice temperature is 452 K ($\sigma$  = 1.4). The amplitude and wavelength are $\text{F}_{0}$ = 0.025 a.u. and 1.5 $\mu m$ respectively. A uniformly trapezoidal envelope laser with two rising, two falling and eight plat-top cycles is used in all simulations. The cutoff frequency is denoted by the vertically line. }
		\label{Fig2}
	\end{figure}
	
The electronic structure of solids under certain temperature is obtained from the diagonalization of the time-independent Hamiltonian $\hat{\text{H}}_0 = \frac{{{{\hat p}^2}}}{{2m}} + V\left( x \right)$.  The calculated details of  electronic structure could be found in Appendix C. We adopt the all valence band states as the initial states for each structural configuration. And for each occupied state $|\psi_{n}\rangle$ we independently solve the time-dependent Schr\"odinger equation (TDSE) under the velocity gauge as 
	\begin{equation}\label{Eq2}
		i \frac{\partial }{{\partial t}}|{\psi_n}(t)\rangle  = \hat{\text{H}}(t)|{\psi_n}(t)\rangle,
	\end{equation}
and then calculate time-dependent current by ${j}(t) = -\sum_{n}\langle {\psi_n}(t)|\hat{\text{p}} + \text{A}(t)|{\psi_n}(t)\rangle$. The laser vector potential with frequency $\omega$ is A(t) =  -$\int_{-\infty}^t \text{F}(t')dt'$, and F(t) is the electric field. $\text{A}_{0}$ and $\text{F}_{0}$ are their respective amplitudes.

The duration of femtosecond (fs) laser pulses irradiating targets is comparable to the timescale of phonon modes with the 100-fs magnitude. Furthermore, the laser pulses with a kilohertz (KHz) repetition frequency \cite{luu} could encounter thousands of atomic-chain arrangements.  Different configurations have a same structural fluctuation but posses the various spacial arrangements at a certain temperature. The events occurring in normal distribution generator with a given fluctuation $\sigma$ produce the atomic-chain configurations. The experimental HHG spectrum contains thousands of currents contributed by different structural configurations. To obtain the final harmonic spectrum, we coherently sum the $j(t)$ of each correlative structural configuration as J(t) = $\sum j(t)$ before the Fourier transform of the total current J(t).
	\begin{figure}[t]
		\includegraphics[width=8 cm,height =7 cm ]{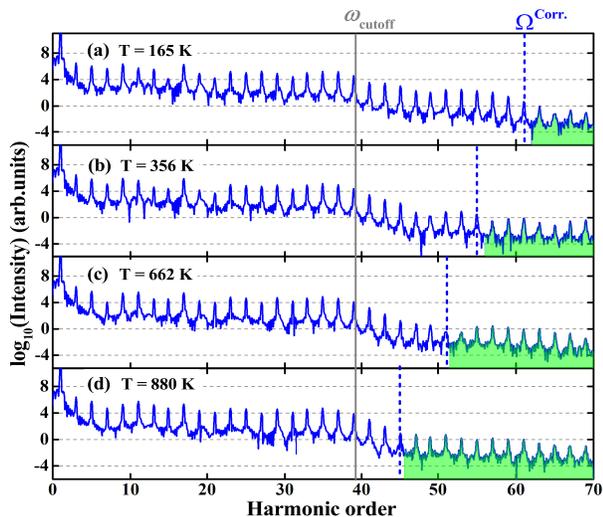}
		\caption{Temperature-dependent HHG spectra. The spatial fluctuations adopted in (a), (b), (c) and (d) are 0.8, 1.2, 2 and 4 a.u. respectively.   } 	\label{Fig3}		
	\end{figure}
	
\section{Discussions}
We firstly present the HHG spectra with the increasing configuration number. Taking $\sigma$  = 1.4 a.u. as an example,  Figs. \ref{Fig2}(a-c) respectively show HHG spectra contributed by the 1, 100 and 4000 configurations. In Fig. \ref{Fig2}(a), the HHG spectrum only contains the odd-order harmonics below 41th harmonic and the even-order harmonics then appear in the spectral region beyond the cutoff frequency, as shown by the shadow zone. As more structural configurations are adopted in Figs. \ref{Fig2}(b-c), the signals of the even-order harmonics are suppressed, but those of odd-order harmonics are remained. Moreover, the plateau-zone intensities do not change, but the harmonic intensities for the spectroscopic region beyond the robust cutoff frequency drop about three orders of magnitude. Statistics of configurations imply dephasing process, which leads to the shrink of even-order harmonic zone.  Keeping the each configuration corresponding to one of atomic arrangements in mind, each laser pulse will capture one configuration of the vibrating structure, in which the translation invariance of atomic spacing has been broken. Thus, one observes the even-order harmonics beyond the cutoff frequency. The disappearance of the even-order harmonics in the Fig. \ref{Fig2}(c) indicates that the  atomic chain will retrieve the intrinsic and periodic symmetry with sufficient configuration number. In brief, the vibration-induced disorder could not give rise to emergence of even-order harmonic in solid-state HHG spectra, which reaches consensus with the experimental observations \cite{bai_ya,luu}. 
	
	\begin{figure}[t]
		\includegraphics[width=8cm,height = 7cm ]{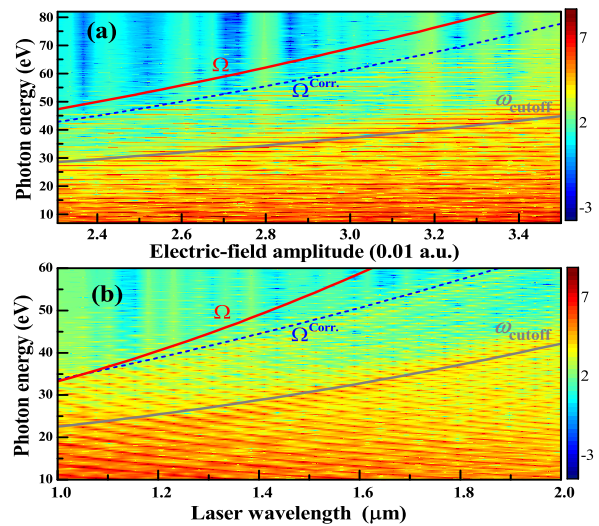}
		\caption{Dependencies of cutoff frequency and maximal photon energy on the laser parameters. The HHG spectra vary with  (a) electric field amplitude and (b) wavelength under the lattice temperature about 100 K ($\sigma$ = 0.5 a.u.). The solid and dashed curves are predicted by their respective formulas. } 	\label{Fig4}		
	\end{figure}

Then, we discuss the impact of lattice temperature on the cutoff frequency and the maximum photon energy in the HHG spectra. Note that all the results discussed below are obtained from the number of 4000 configurations. As the temperature increases, one sees the robustness of cutoff frequency and the decrease of the maximum photon energy, as shown in Figs. \ref{Fig3}(a-d). Here the maximal photon energies marked by blue dashed lines in  Figs. \ref{Fig3}(a-d) are confirmed by performing time-frequency analyses on the spectral region beyond the cutoff frequency, as presented in Fig. \ref{FigS3} of Appendix D. The dephasing of quantum trajectory is enhancing with growing lattice temperature, thus lowing the maximal photon energy.

The cutoff frequency $\omega_{\text{cutoff}}$ of the HHG plateau, in fact, is determined by the maximal kinetic energy obtained by the electrons when the distance between electron and hole satisfies the condition of zero displacement (i.e. closed-trajectory model). As shown by the Figs. \ref{FigS2}(a-d) of Appendix C, we conclude that the nuclear vibration under the finite temperature just slightly perturbs the energy band structure, which clarifies the robustness of cutoff frequency in Figs. \ref{Fig3}(a-d). This robust cutoff energy is depicted as $\omega_{\text{cutoff}} = I_p + 3.17U_p$, where $I_p$ is the ionized potential of solids (about 16 eV) and the ponderomotive energy, $U_p$, is given by $\text{F}_{0}^2/4\omega^2$.
	
	\begin{figure}[t]
		\includegraphics[width=9 cm,height = 8 cm ]{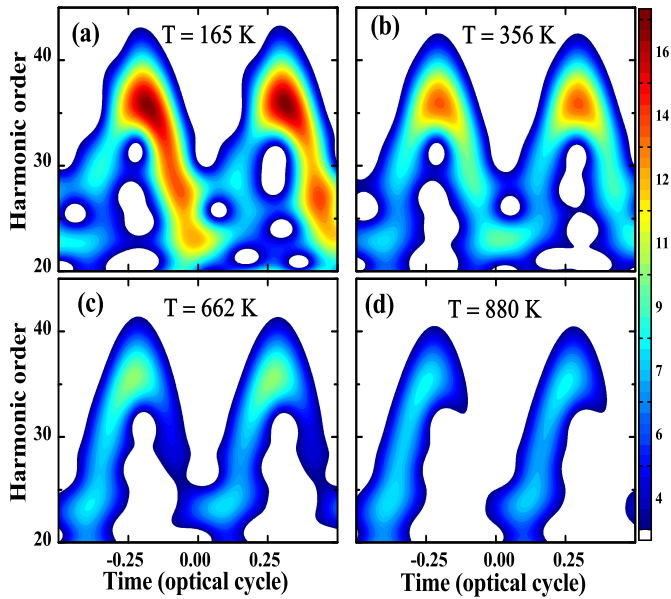}
		\caption{Time-frequency analyses of the HHG spectra in Fig. \ref{Fig3}. } 	\label{Fig5}		
	\end{figure}
	
However, the maximal photon energy in HHG spectrum is related to the coherent overlap between electron and hole wave packets. Taking into account the wave-like property of electron wave packet in solids, one could remove the condition of zero displacement in recombined step, i.e. open-trajectory model \cite{dephasing,Crosse}. When the coherent width ($\mathcal{D}$) between electron and hole wave packets  is greater than the excursion distance of electron, one obtains the high-harmonic photon energy formulated as $\Omega$  = $I_p$ + 2$U_p [\sin(\omega t) - \sin(\omega t') ]^2$ under the free-scattering situation \cite{Corkum,Lewenstein}, where $t'$ and $t$ are the ionized and recombined times respectively. The maximal photon energy occurs for $\omega t'$ = $\pi$/2 and $\omega t$ = 3$\pi$/2, which results in the upper limit of the emitted photon energy $\Omega$ = $I_p$  + 8$U_p$. Keeping the wave-particle duality in mind, dephasing of electron wave packet is characterized by its particle-like property. The coherent width could be written as $\mathcal{D}$ = $2\text{A}_{0}\text{T}_2$, where $\text{T}_2$ is the dephasing time. The formula derivation could be found in the Appendix E. The HHG induced by quantum coherence will involve an electron-hole-pair polarization energy depicted as $\text{F}_{0}\mathcal{D}$. Thus the emitted maximal photon energy could be corrected as $\Omega^{\text{corr.}}$ = $\omega_{\text{cutoff}}$ +  $\text{F}_{0}\mathcal{D}$ under the situation of dephsing.

To demonstrate the formulas for the cutoff energy ($\omega_{\text{cutoff}}$) and the maximal photon energies ($\Omega$ and $\Omega^{\text{corr.}}$), in Figs. \ref{Fig4}(a) and \ref{Fig4}(b) we respectively present the dependencies of HHG spectra on the electric amplitude and wavelength of laser pulses. The emitted photon energies predicted by three formulas reach a good agreement with the results of TDSE simulations in both Fig. \ref{Fig3} and Fig. \ref{Fig4}. Role of the lattice temperature on the maximal photon energy has been attributed to temperature-induced dephasing in $\Omega^{\text{corr.}}$. To further confirm the dephasing process induced by the finite-temperature nuclear vibrations,  we perform the temporal profiles of HHGs. In Figs. \ref{Fig5}(a-d) we display the time-frequency analyses of the HHG spectra in Figs. \ref{Fig3}(a-d), respectively. In Fig. \ref{Fig5}(a) one could observe that the spectroscopic intensities of the long trajectories are stronger than those of the short trajectories. However, this situation will be reversed with the growing lattice temperature in Figs. \ref{Fig5}(b-d), and the dephasing rate of long trajectory is greater than that of short trajectory. Similarly, this tendency is also observed in the quantum trajectories of the spectroscopic zone beyond cutoff frequency, as shown in Fig. \ref{FigS3}(b). The laser-driven electronic wave packets undergo the non-negligible dissipation induced by the electron-phonon scattering. Furthermore, the long-trajectory electrons will travel a longer excursion time than those of the short trajectory, which causes the dissipation on long trajectory is dominated than that of short trajectory.

	\begin{figure}[t]
		\includegraphics[width=8cm ,height = 6  cm ]{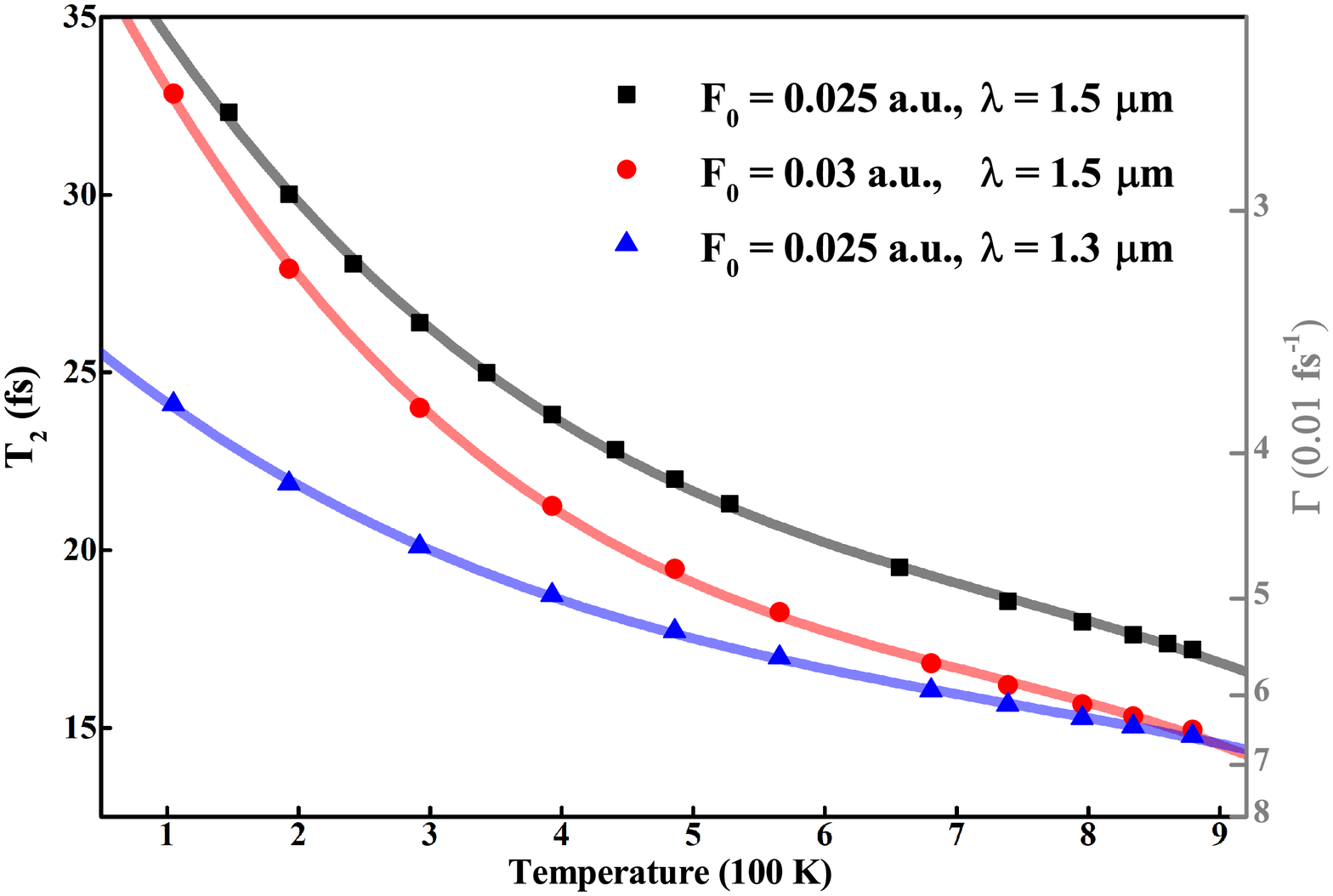}
		\caption{The extracted dephasing times ${\text{T}_2}$ with varying  lattice temperatures and laser parameters are marked with square, circular and triangle dots. The dephasing times predicted by formula are plotted as solid curve. The parameters $\beta$ for the black, red and blue lines are adopted respectively -0.19, -0.24 and -0.16 fs$^{-1}$, which may be influenced by the laser-field-dependent ionized rate. }
		\label{Fig6}
	\end{figure}

The single active electron approximation suffers from neglecting the dephasing from the inelastic scattering by lattices. To remedy the dephasing in the single active electron approximation, we incorporate an imaginary potential $iU(x)$ into the field-free Hamiltonian, which is denoted as  ${\hat H_0} - iU(x) $ \cite{dephasing,Du1}. In such a treatment, the eigenenergy becomes complex value and its imaginary part $i\Gamma$ corresponds to the depahsing term.

We take the HHG spectra obtained from the fixed-nuclei periodic chain including phenomenological dephasing as theoretical input. The temperature-dependent HHG spectra are regarded as experimental results. To extract the dephasing time, we then perform an analysis as done in typical comparisons of theory and experiment \cite{solid2,solid3,propagation}. Based on the fact that the relative height of the HHG plateau depends very sensitively on the dephasing rate, in the theoretical calculations the value of  $\Gamma$ is used as a freely adjustable parameter of the dephasing HHG spectra.  As one can see in Fig. \ref{FigS4} of Appendix F, we obtain rather good agreement between temperature-dependent spectra and dephasing spectra by systematically decreasing the used dephasing times, along with increasing temperature. By comparing delicately the finite-temperature HHG spectrum with that obtained from the theoretical simulations, we could extract the dephasing time with varying lattice temperatures and laser parameters, as presented by the dots in Fig. \ref{Fig6}. To understanding these results, we introduce a formula for the finite-temperature dephasing rate as 
\begin{equation}\label{Eq3}
		\Gamma = \gamma \frac{L_{max}}{a-\sigma} + \beta,
\end{equation}
in which the electronic maximal classical excursion $L_{max} =  \frac{\pi \text{F}_{0}}{\omega^2}$ (cf. Appendix E), the ratio of $L_{max}$ and effective lattice constant $a-\sigma$ denotes the collision times between electrons and lattices. The parameter $\gamma$ adopted as 0.026 fs$^{-1}$ represents the contributed dephasing rate from each electron-lattice collision and is relative to type of solid-state materials. One can observe that the curves predicted by Eq. (\ref{Eq3}) conform well with the extracted dephasing times ($\text{T}_{2} = \frac{1}{2\Gamma}$) marked with dots in the Fig. \ref{Fig6}. Therefore, the dephasing time can be quantitatively linked with the lattice temperature.

\section{Conclusion} 
To summarize, mechanism about the temperature-induced dephasing in the solid-state HHG has been revealed. We develop a temperature dependent lattice vibration model by characterizing the spacing fluctuations between atomic pairs. Moreover, the lattice translation symmetry is not broken by the vibration-induced fluctuation, which is verified by the missing even-order harmonics in this work as well as its experimental observations. In the temperature-dependent HHG spectra, we clarify the impact of temperature-induced depasing on the maximal photon energy under the open-trajectory model. Finally a formula shedding new light on the timescale of electron-lattice scattering is introduced and also confirmed by the temperature-dependent HHG spectra. Our results reaffirm the significance of quantum decoherence in the quantitative analysis of HHG experiments.

	\section*{Acknowledgment}
	This work is supported by the National Natural Science Foundation of China (NSFC) (Grant No. 11904331).
	
	T.-Y.D. and C.M. contributed equally to this work.

\section*{APPENDIX A: Mean-square relative displacements in the temperature dependent lattice vibration model}
The peak of atomic pair distribution function (PDF)  in simple crystals can be described approximately by the Gaussian-type function with a fluctuation $\sigma_{ij}$. By projecting onto the concerning vector linking the atom pairs, the mean-square relative displacement of atom pairs, is denoted as
	\renewcommand\theequation{A1}
	\begin{equation}
		\sigma^2_{ij} = \langle [({\bf u_i - u_j})\cdot {\bf \hat e_{ij}}]^2 \rangle,
	\end{equation}
where ${\bf u_i, u_j}$ are thermal displacements of atoms $i$ and $j$ from their equilibrium positions. The vector ${\bf \hat e_{ij}}$ is an unit vector along the direction connecting the atoms $i, j$, and the angular brackets $\langle \rangle$ indicate an ensemble average. This equation can be expanded as
	\renewcommand\theequation{A2}
	\begin{equation}
		\sigma^2_{ij} = \langle [{\bf u_i \cdot \hat e_{ij}}]^2 \rangle
		+ \langle [{\bf u_j \cdot \hat e_{ij}}]^2 \rangle
		-2\langle ({\bf u_i \cdot \hat e_{ij}})
		({\bf u_j \cdot \hat e_{ij}}) \rangle .
		\label{eq;msrd2}
	\end{equation}
	Here the first two terms correspond to the mean-square thermal displacement of atoms {\it i} and {\it j}. The third term is a displacement correlation function, which encodes information about the atomic motional correlations. For crystals constituted by monatomic type, the $\sigma^2_{ij}$ is expressed in terms of the lattice phonons as follows, 
		\renewcommand\theequation{A3}
	\begin{equation}
		\sigma^2_{ij}
		= {2  \over NM}\,\sum_{{\bf k}, s} {{({\bf \hat e_{k,s}}
				\cdot {\bf \hat e_{ij}})^2 \over \omega_s({\bf k})} }
		[n(\omega_s({\bf k}))+{1 \over 2}][1-\cos(\bf {k \cdot r_{ij}})],
		\label{eq;msrd3}
	\end{equation}
	where $\omega_s({\bf k})$ is a phonon-mode frequency with crystal wave vector ${\bf k}$ in branch $s$, $n(\omega_s({\bf k}))$ is the phonon occupation number, ${\bf \hat e_{k,s}}$ is the polarization vector of the ${\bf k}, s$ phonon mode, $N$ is the number of atoms and $M$ is the mass of an atom.  $\bf r_{ij}$ is the relative interval between atoms {\it i} and {\it j} along the vector ${\bf \hat e_{ij}}$.  In calculation the phonon frequency ($\omega_s({\bf k})$) and polarization vector ($\bf \hat{e}_{k,s}$) could be obtain by solving the dynamical matrix using up to the high-order nearest-neighbor interatomic force parameters. Note that the force parameters of crystals become extractable by the open-source first-principle codes such as {\it Phonopy} and {\it Quantum Espresso}. And then the phonon dispersion curves can be obtained by using the Born von-Karman (BvK) model.
	As mentioned above, the force constants must be known in advance to obtain all phonon modes via the BvK model calculation. 
	
	Then, we will simplify the result in Eq. (\ref{eq;msrd3}) using some approximations to describe the effects of the lattice vibrations on the peaks of  PDF without knowing the force constants. 
	Following the works reported by Debye \cite{Debye}, and Beni and Platzmann \cite{Beni}, one could make no distinction between longitudinal and transverse phonon branches and further take account of a spherical average. Then Eq. (\ref{eq;msrd3}) reduces to
		\renewcommand\theequation{A4}
	\begin{equation}
		\sigma^2_{ij}
		= \biggl < {2 \over M \omega}
		\Bigl [n(\omega)+{1\over 2}\Bigr]\Bigl[1-\cos(\bf {k \cdot
			r_{ij}})\Bigr] \biggr >,
		\label{eq;msrdave}
	\end{equation}
	where $\langle \cdot \cdot \cdot \rangle$ is the average over the
	3$N$ branches and $N$ is the number of atoms. This equation is a general expression for all crystal materials and is independent of the number of atoms per unit cell. Using the Debye approximation, $\omega = ck$, we can rewrite Eq. (\ref{eq;msrdave}) as
	follows \cite{Sevillano}:
		\renewcommand\theequation{A5}
	\begin{equation}
		\sigma^2_{ij}
		= {2 \over 3NM}\,\int_{0}^{\omega_D}  d\omega \,{\rho (\omega) \over
			\omega} \biggl [n(\omega)+{1\over 2} \biggr]\biggl[1-{\sin(\omega
			r_{ij}/c) \over \omega r_{ij}/c} \biggr],
		\label{eq;msrddw}
	\end{equation}
	where $\rho (\omega) = 3N (3 \omega^2/ {\omega_D}^3) $ is the phonon density of states, $c$ is the sound velocity,  n($\omega$) is the phonon occupation number. $\omega_D ~{= c k_D}$ is the Debye cutoff frequency. The Debye wave vector is given by $k_D = (6\pi^2N/V)^{1/3}$ where $N/V$ is the number density of atom in the crystal. After integrating over $\omega$, we obtain
		\renewcommand\theequation{A6}
	\begin{eqnarray}
		\sigma^2_{ij} &=& {6 \over M
			\omega_D}\biggl[\frac{1}{4}+\biggl(\frac{T}{\Theta_D}\biggr)^2
		\Phi_1 \biggr] - {6 \hbar \over M \omega_D} \biggl[ {1-\cos(k_D
			r_{ij}) \over {2(k_D r_{ij})^2}}
		\nonumber \\
		&+& \biggl(\frac{T}{\Theta_D}\biggr)^2 \int_{0}^{\frac{\Theta_D}
			T} \frac{\sin(\frac{k_D r_{ij} T x}{\Theta_D})/(\frac{k_D r_{ij}
				T}{\Theta_D})} {e^x-1}~dx \biggr], \label{eq;msrd4}
		%\end{equation}
	\end{eqnarray}
	where $\Phi_1 = \int_{0}^{\Theta_D/T} x (e^x-1)^{-1}~dx$, $x$ is a
	dimensionless integration variable and $\Theta_D$ (=$\hbar
	\omega_D/k_B$) is the Debye temperature. Here, the first term corresponds to the usual uncorrelated mean-square thermal displacements (2$\langle u^2 \rangle$) and the second term is the displacement correlation function. This result is known as the correlated Debye model \cite{Beni,Bohmer,Sevillano}.
	
	For simplicity, we could derive an uncorrelated Debye model by neglecting the third term (a displacement correlation function) in Eq. (\ref{eq;msrd2}). Only the first term in Eq. (\ref{eq;msrd4}) could be retained, and the temperature-dependent mean-square relative displacement of atom pairs {\it i} and {\it j} are finally denoted as 
		\renewcommand\theequation{A7}
	\begin{equation}
		\sigma_{ij}^2 =  {6 \over M
			\omega_D}\biggl[\frac{1}{4}+\biggl(\frac{T}{\Theta_D}\biggr)^2
		\Phi_1 \biggr].
		\label{eq;msrd5}
	\end{equation}

	\section*{APPENDIX B:Corresponding relationship between fluctuation $\sigma$ and lattice temperature}
\begin{figure}[t]
		\includegraphics[width= 7  cm,,height=6 cm  ]{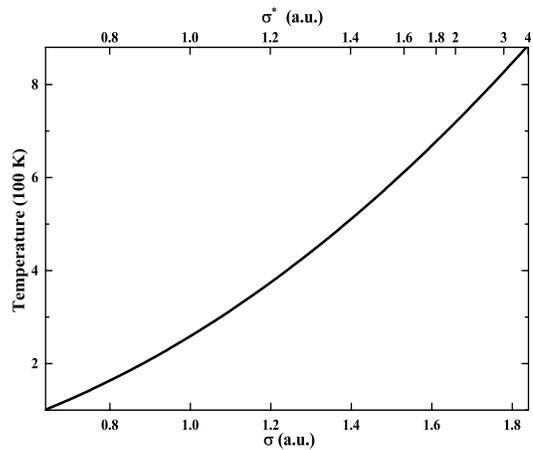}
		\caption{Lattice-vibration-induced spatial fluctuation $\sigma$ as a function of temperature according to Eq. (\ref{eq;msrd5}). The spacing fluctuation $\sigma$ is defined by including the role of truncation in the effective fluctuation $\sigma^{*}$. }
		\label{FigS1}
	\end{figure}

	In Fig. \ref{FigS1} the spacing fluctuation $\sigma$ is obtained from the full width at half maximum (FWHM) of effective fluctuation $\sigma^{*}$ in the truncated normal distribution. Taking into Debye temperature (150 K) and carbon atomic mass into account, the fluctuation of vibration amplitude is linked with the lattice temperature clarified by Eq. (\ref{eq;msrd5}), as presented in Fig. \ref{FigS1}. For a certain material, ${\Theta _D}$ can be used as an empirical parameter by measurements of the specific heat or channeling experiments. Finally, the atomic-chain structure of real system under different temperatures can also be signed by the structure factor $\sigma$.

		\begin{figure}[t]
		\includegraphics[width=8.6 cm,height=8 cm ]{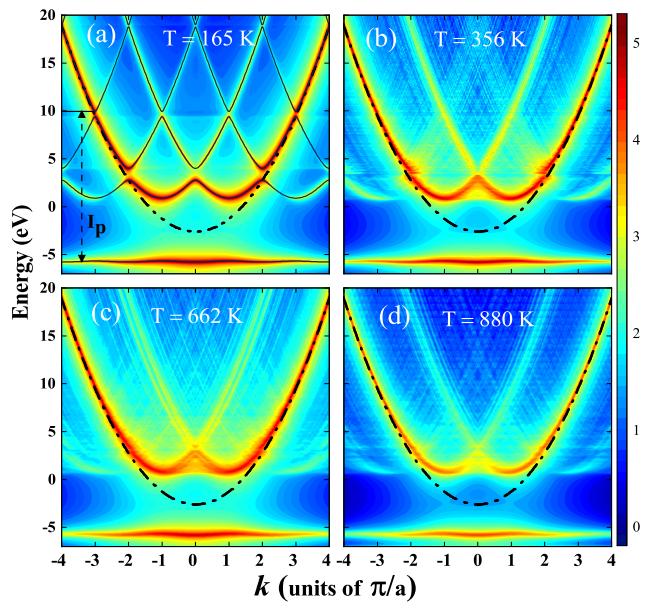}
		\caption{The electronic structure under various lattice temperatures. In (a-d) the results respectively correspond to the spatial fluctuation $\sigma$ = 0.8, 1.2, 2 and 4 a.u.  In (a) the black solid curves are the zeros-temperature ($\sigma$ = 0) energy bands. The black dash-dotted parabolic curve in (a-d) is similar to the energy band for a free electron. According to the discrepancy between zero-temperature energy bands and parabolic band, we define the ionization potential (I$_p$) of crystals, as denoted in (a). }
		\label{FigS2}
	\end{figure}
	
	\section*{APPENDIX C:Temperature-dependent electronic structure}
	To generate a smoothly varying model potential, the atomic-chain potential is defined as 
	\renewcommand\theequation{C1}
	\begin{equation} 
		\text{V}(x) =  - {\text{V}_0} \sum_{j}  \left\{ {\exp [ - \frac{{{{(x - {x_j})}^2}}}{{2{\alpha ^2}}}] + \exp [ - \frac{{{{(x - {x_{j + 1}})}^2}}}{{2{\alpha ^2}}}]} \right\}
	\end{equation}
	where V$_0$ = 0.52 a.u. and $\alpha$  = 0.08 a.u., and the number of atoms in the vibrating-lattice chain adopts N = 500. 
	We utilize spectral function technique to directly analyze electronic structure from eigen-wavefunction in coordinate representation. The function is read as
	\renewcommand\theequation{C2}
	\begin{equation} 
		I(k,E) = \sum\limits_n {\delta (E - {E_n})} |\langle k|{\psi _n}\rangle {|^2}
	\end{equation}
	where ${E_n}$, $|{\psi _n}\rangle$ are respectively the eigenenergy and eigen-wavefunction by diagonalization of time-independent Hamilton ${\hat H_0}$, $|k\rangle$ is plane wavefunction ${e^{ikx}}$.  The energy bands with various $\sigma$ hold the similar energy band structure, as shown by Figs. \ref{FigS2}(a-d). As $\sigma$ increases, in reduced Brillouin zone many bands disappear due to the destruction of spatial symmetry. However, the low energy band and parabolic band are maintained well, as presented by the black dash-dotted parabolic curves in Figs. \ref{FigS2}(a-d). Therefore, temperature-induced lattice vibration will not destroy the energy band but only perturb the electronic structure. In Fig. \ref{FigS2}(a), one sees that the low-temperature energy bands (colormap) have a great agreement with zero-temperature energy bands (black solid curves), in which the temperature-dependent energy bands (colormap) in extended Brillouin zone follow the parabolic dispersion (black dotted-dash curve) when the eigenenergies are larger than 10 eV. Here the ionization potential (I$_p$) of crystals is defined as 16 eV. Another impact of the growing lattice temperature on the energy band is that the small energy gap is gradually closed.

	\section*{APPENDIX D:Effect of lattice temperature on the maximal photon energy.}
\begin{figure}[t]
		\includegraphics[width=9 cm ]{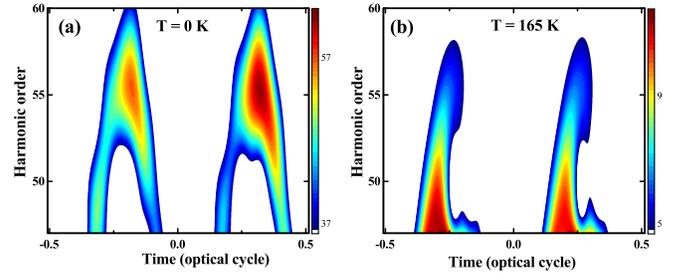}
		\caption{Time-frequency analysis for the HHG spectra beyond the cutoff frequency under two temperatures. The laser parameters are same as Fig. \ref{Fig3}.}
		\label{FigS3}
	\end{figure}

	The yields of HHG spectra beyond the cutoff frequency decay fast with growing $\sigma$. We perform a time-frequency analyses for the HHG spectra beyond the cutoff frequency. The long and short trajectories are complete under the case of zeros temperature ($\sigma$ = 0), as shown in Fig. \ref{FigS3}(a). However, in Fig. \ref{FigS3}(b) with lattice temperature 165 K the long trajectory disappears and the emitted maximal photon energy is decreased.	

\section*{APPENDIX E:Maximal classical excursion and coherent width}
The electron-hole pair wave function can be denoted as 
	\renewcommand\theequation{E1}
\begin{equation}\label{A1}
	P(\text{k},t) = -\int_{0}^{\infty} \ d\tau \ e^{iS(\text{k},t,\tau) - i\omega t } \zeta, 
\end{equation}
with the classical action given by 
	\renewcommand\theequation{E2}
\begin{equation}\label{A2}
	\begin{split}
		& S(\text{k},t,\tau)   =  - \int_{t-\tau}^{t} \  dt' \  \Delta \text{E}(\text{k},t') + i\Gamma \tau + \omega t, \\
		& =  - \int_{t-\tau}^{t} \  dt' \ \lbrace 2\text{U}_{p} [\frac{\text{k}}{\text{A}_{0}} - \sin (\omega t') ]^2  \rbrace  + i\Gamma \tau + \omega t.
	\end{split}
\end{equation}
Here the dephasing induced by our concerned lattice vibration is introduced via the dephasing rate $\Gamma$. $\zeta$ is the Rabi frequency of multiplying the electric field amplitude and transition dipole moment, and $\Delta \text{E} = \frac{\text{k}^{2}}{2m_{R}}$ is the energy dispersion of electron-hole pair under the multi-photon resonant excitation  and parabolic approximations, and $m_{R}$ is the electron-hole reduced mass.  $\text{k}(t) = \text{k}_{0} -  \text{A}_{0}\sin(\omega t)$ is the canonical momentum of electron-hole pair, and $\text{U}_{p} = \text{A}^{2}_{0}/4m_{R}$ is the ponderomotive energy. 
We firstly expand the Eq. (\ref{A2}) and neglect the indirect driving term with 2$\omega$ frequency component. Then we obtain the saddle-point equation by the first derivative of classical action $S$ with respect to the recollision time $t$, which can be written as 
	\renewcommand\theequation{E3}
\begin{equation}\label{A3}
	\omega_s = 2\text{U}_{p} \{\sin[\omega (t-\tau)] - \sin(\omega t)  \}^{2}.
\end{equation}
Finally, the highest energy can be obtained at the condition for $\omega t = 3\pi/2$ and $\omega \tau = \pi$. 
For trajectories that maximizes Eq. (\ref{A3}) ($\omega t = 3\pi/2$ and $\omega \tau = \pi$), the electronic maximal classical excursion is 
	\renewcommand\theequation{E4}
\begin{equation}\label{A5}
	L_{max} = - \int^{t}_{t-\tau}
	\frac{\partial\Delta \text{E}}{\partial \text{k}}\bigg|_{\text{k}(t')} dt' =  \frac{\pi \text{F}_{0}}{m_{R}\omega^2}
\end{equation}

\begin{figure}[t]
	\includegraphics[width= 8cm,height = 7cm ]{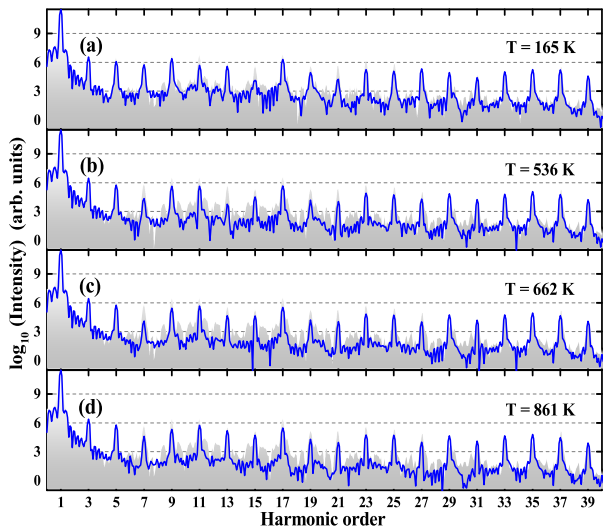}
	\caption{Comparisons between theoretical (gray background curves) and experimental (blue solid curves) examples of HHG spectrum. }
	\label{FigS4}
\end{figure}

To obtain an estimate of the wave function diffusion width, we expand the trigonometric functions to the first order in $t-\tau$ where the points of highest energy electron-hole pair trajectories in Eq. (\ref{A3}) are included only. Evaluating the excursion time $\tau$ integral in Eq. (\ref{A1}) approximatively gives rise to the wave function distribution in momentum space, which can denoted as
	\renewcommand\theequation{E5}
\begin{equation}\label{A4}
	P(\text{k},t) \approx  \frac{ e^{iS(\text{k},t) - i\omega t } \zeta}{2U_{p}(\frac{\text{k}^2}{\text{A}^2_0}  +  2\frac{\text{k}}{\text{A}_0})-i\Gamma }.
\end{equation}
Hence the half maximum of wave function in the crystal-momentum space appears at $\text{k}_{1/2}$ = $-\text{A}_{0}(1\pm \sqrt{1+\Gamma/2\text{U}_{p}}  )$ and then one achieves a full-half-maximum of  $\Delta \text{k}_{\text{FWHM}}$ = $\text{A}_{0}\Gamma/2\text{U}_p$. The coherent width ($2/ \Delta \text{k}_{\text{FWHM}}$) between electron and hole wave packets is given as $\mathcal{D}$ = $2\text{A}_{0}\text{T}_2/m_{R}$, where the dephasing time is clarified as ${\text{T}_2} = \frac{1}{{2\Gamma}}$ and the reduced mass $m_{R}$ adopts 1 a.u. in this work.

\section*{APPENDIX F:Comparisons between theoretical and experimental examples of HHG spectrum.}
By including the role of phenomenological dephasing, the HHG spectra obtained from the fixed-nuclei periodic chain are regarded as the theoretical spectra.  The temperature-dependent HHG spectra are the experimental HHG spectra. To extract the dephasing time, we have performed a typical comparisons of theory and experiment. Comparing the theoretical (gray background curves) and experimental (blue solid curves) examples of HHG spectrum, one can observe a great agreement with each other.


\begin{thebibliography}{30}
\bibitem{Chin} A. H. Chin, O. G. Calder\'on, and J. Kono, Phys. Rev. Lett. {\bf 86}, 3292 (2001).
	\bibitem{Krausz} F. Krausz and M. Ivanov, Rev. Mod. Phys. {\bf 81}, 163 (2009).
	\bibitem{solid1} S. Ghimire, A. D. DiChiara, E. Sistrunk, P. Agostini, L. F. DiMauro,  and D. A. Reis,  Nat. Phys.  {\bf 7}, 138 (2011).
	\bibitem{solid2} O. Schubert, M. Hohenleutner, F. Langer, B. Urbanek, C.
	Lange, U. Huttner, D. Golde, T. Meier, M. Kira, S. W. Koch, and R. Huber, Nat. Photon. {\bf 8}, 119 (2014).
	\bibitem{solid3} M. Hohenleutner, F. Langer, O. Schubert, M. Knorr, U. Huttner,
	S. W. Koch, M. Kira, and R. Huber, Nature {\bf 523}, 572 (2015).
	\bibitem{Du2} T.-Y. Du, Opt. Lett. {\bf 46}, 2007 (2021).
	\bibitem{solid4} T. T. Luu, M. Garg, S. Y. Kruchinin, A. Moulet, M. T. Hassan,
	and E. Goulielmakis, Nature {\bf 521}, 498 (2015).
	\bibitem{solid5} G. Ndabashimiye, S. Ghimire, M. Wu, D. A. Browne, K. J. Schafer, M. B. Gaarde, and D. A. Reis, Nature {\bf 534}, 520 (2016).
	\bibitem{solid6} G. Vampa, T. J. Hammond, N. Thir\'e, B. E. Schmidt, F. L\'egar\'e, C. R. McDonald, T. Brabec, and P. B. Corkum, Nature {\bf 522}, 462 (2015).
	\bibitem{solid7} Y. S. You, D. Reis, and S. Ghimire, Nat. Phys. {\bf 13}, 345 (2017).
 	\bibitem{Du3} T.-Y. Du and S.-J. Ding, Phys. Rev. A {\bf 99}, 033406 (2019).
	\bibitem{solid8} N. Yoshikawa, T. Tamaya, and K. Tanaka, Science {\bf 356}, 736 (2017).
	\bibitem{Du4} T.-Y. Du, Phys. Rev. A {\bf 100}, 053401 (2019). 
	\bibitem{graphene1} S. A. Mikhailov and K. Ziegler, J. Phys.: Condens. Matter {\bf 20}, 384204 (2008).
	\bibitem{graphene2} S. A. Mikhailov, Phys. Rev. B {\bf 93}, 085403 (2016).


	
	\bibitem{topo1} D. Bauer and K. K. Hansen, Phys. Rev. Lett. {\bf 120}, 177401 (2018).
	\bibitem{topo2} C. J{\"u}r{$\ss$} and D. Bauer, Phys. Rev. B {\bf 99}, 195428 (2019).
	\bibitem{mott1} Y. Murakami, M. Eckstein, and P. Werner, Phys. Rev. Lett. {\bf 121}, 057405 (2018).
	\bibitem{mott2} Y. Murakami and P. Werner, Phys. Rev. B {\bf 98}, 075102 (2018).
	
	\bibitem{disorder1} K. Chinzei and T. N. Ikeda, Phys. Rev. Res. {\bf 2}, 013033 (2020).
	\bibitem{disorder2} C. Yu, K. K. Hansen, and L. B. Madsen, Phys. Rev. A {\bf 99}, 013435 (2019). 
	\bibitem{disorder3} G. Orlando, C.-M. Wang, T.-S. Ho, and S.-I. Chu,  J. Opt. Soc. Am. B {\bf 35}, 680 (2018).
	\bibitem{disorder4} G. Orlando, T.-S. Ho, and S.-I. Chu, J. Opt. Soc. Am. B {\bf 37}, 1540 (2020).
	\bibitem{luu} T. T. Luu,  Z. Yin, A. Jain, et al.  Nat. Commun. {\bf 9}, 3723 (2018). 
	
	\bibitem{Uchida} K. Uchida, G. Mattoni, S. Yonezawa, F. Nakamura, Y. Maeno, and K. Tanaka, Phys. Rev. Lett. {\bf 128},  127401 (2022).
	\bibitem{Heissler} P. Heissler, E. Lugovoy, R. H\"orlein, L. Waldecker, J. Wenz, M. Heigoldt, K. Khrennikov, S. Karsch, F. Krausz, B. Abel, and G. D. Tsakiris, New J. Phys. {\bf 16} 113045 (2014).
	
	\bibitem{JMziman} J. M. Ziman, Models of Disorder (Cambridge University Press, Cambridge, England, 1979).
	\bibitem{RDF1} T. Shiga, T. Murakami, T. Hori, O. Delaire, and J. Shiomi, Appl. Phys. Express {\bf 7}, 041801 (2014).
	\bibitem{Debye} P. Debye, Ann. Phys. (Leipzig) {\bf 39}, 789 (1912).
	\bibitem{RDF2} I. K. Jeong, R. H. Heffner, M. J. Graf, and S. J. L. Billinge, Phys. Rev. B  {\bf 67}, 104301 (2003).
	
	
%	\bibitem{sm} See Supplemental Material for additional information.
	
	
	\bibitem{x-ray} D. Schlesinger, K. Thor Wikfeldt, L. B. Skinner,  C. J. Benmore, A. Nilsson, and L. G. M. Pettersson, J. Chem. Phys. {\bf 145}, 084503 (2016).
	\bibitem{neutron} B. T. M. Willis and C. J. Carlile, Experimental Neutron Scattering (Oxford University Press, Oxford, 2009).
	\bibitem{electron} R. F. Egerton, Electron Energy Loss Spectroscopy in the Electron Microscope (Plenum, New York, 1986).
	
	\bibitem{Urquidi} J. Urquidi, S. Singh, C. H. Cho, and G. W. Robinson, Phys. Rev. Lett. {\bf 83}, 2348 (1999).
	\bibitem{Ojovan} M. I. Ojovan and M. I. Ojovan, J. Phys. Chem. B {\bf 124}, 3186 (2020).
	\bibitem{zengaiwu} A.-W. Zeng and X.-B. Bian, Phys. Rev. Lett. {\bf 124}, 203901 (2020).
	\bibitem{bai_ya} Y. Bai, F. Fei, S. Wang, N. Li, X. Li, F. Song, R. Li, Z. Xu and P. Liu, Nat. Phys. {\bf 17}, 311 (2021).
	\bibitem{Crosse} J. A. Crosse and R.-B. Liu, Phys. Rev. B {\bf 89}, 121202(R) (2014).


	\bibitem{dephasing} G. Wang and T.-Y. Du, Phys. Rev. A {\bf 103}, 063109 (2021).
	
	\bibitem{Corkum} P. B. Corkum, Phys. Rev. Lett. {\bf 71}, 1994 (1993).
	\bibitem{Lewenstein} M. Lewenstein, P. Balcou, M. Y. Ivanov, A. L'Huillier, and P. B. Corkum, Phys. Rev. A {\bf 49}, 2117 (1994).
\bibitem{Du1} T.-Y. Du, Phys. Rev. A {\bf 104}, 063110 (2021).

	\bibitem{propagation} I. Kilen, M. Kolesik, J. Hader, J. V. Moloney, U. Huttner, M. K. Hagen, and S. W. Koch, Phys. Rev. Lett. {\bf 125}, 083901 (2020).
	
	
	
% \bibitem{Debye} P. Debye, Ann. Phys. (Leipzig) {\bf 39}, 789 (1912).
\bibitem{Beni} G. Beni and P. M. Platzman, Phys. Rev. B {\bf 14}, 1514 (1976).
\bibitem{Sevillano} E. Sevillano, H. Meuth, and J. J. Rehr, Phys. Rev. B {\bf 20}, 4908 (1979).
\bibitem{Bohmer} W. Bohmer and P. Rabe, J. Phys. C {\bf 12}, 2465 (1979).
		
		
		
	\end{thebibliography}
\end{document}